\documentclass[12pt,preprint]{aastex}

\def\linebreak{\hfil\break}

\def\
{\hfil\linebreak}

\def\degree{\ifmmode {^\circ}\else {$^\circ$}\fi}
\def\mum{\ifmmode {\rm \mu {\rm m}}\else $\rm \mu {\rm m}$\fi}
\def\arcsec{\ifmmode ^{\prime \prime}\else $^{\prime \prime}$\fi}

\def\inch{\ifmmode ^{\prime \prime}\else $^{\prime \prime}$\fi}
\def\arcmin{\ifmmode ^{\prime}\else $^{\prime}$\fi}

\def\mearth{M$_\oplus$}
\def\mstar{M$_\star$}

\def\2470{[24]--[70]}

\newbox\grsign \setbox\grsign=\hbox{$>$} \newdimen\grdimen \grdimen=\ht\grsign\newbox\simlessbox \newbox\simgreatbox
\setbox\simgreatbox=\hbox{\raise.5ex\hbox{$>$}\llap
     {\lower.5ex\hbox{$\sim$}}}\ht1=\grdimen\dp1=0pt
\setbox\simlessbox=\hbox{\raise.5ex\hbox{$<$}\llap
     {\lower.5ex\hbox{$\sim$}}}\ht2=\grdimen\dp2=0pt

\begin{document}

\title{Rapid Formation of Icy Super-Earths and the Cores of Gas Giant Planets}
\vskip 7ex
\author{Scott J. Kenyon}
\affil{Smithsonian Astrophysical Observatory,
60 Garden Street, Cambridge, MA 02138} 
\email{e-mail: skenyon@cfa.harvard.edu}

\author{Benjamin C. Bromley}
\affil{Department of Physics, University of Utah, 
201 JFB, Salt Lake City, UT 84112} 
\email{e-mail: bromley@physics.utah.edu}
%
%

\begin{abstract}
We describe a coagulation model that leads to the rapid formation of 
super-Earths and the cores of gas giant planets.  Interaction of 
collision fragments with the gaseous disk is the crucial element of this
model.  The gas entrains small collision fragments, which rapidly settle 
to the disk midplane. Protoplanets accrete the fragments and grow to masses 
$\gtrsim$ 1 \mearth\ in $\sim$ 1~Myr. Our model explains the mass distribution
of planets in the Solar System and predicts that super-Earths form more
frequently than gas giants in low mass disks.

\end{abstract}

\keywords{planetary systems -- solar system: formation -- 
planets and satellites: formation}

\section{INTRODUCTION}

Collisional cascades play a central role in planet formation.  In current
theory, planets grow from collisions and mergers of km-sized planetesimals 
in a gaseous disk.  As planets grow, they stir leftover planetesimals along 
their orbits to high velocities. Eventually, collisions among planetesimals 
produce smaller fragments instead of larger, merged objects. Continued 
stirring leads to a cascade of destructive collisions which grinds the 
leftovers to dust. This process (i) explains the masses of terrestrial 
planets \citep{kb06} and Kuiper belt objects \citep{kbod08} and (ii) produces 
debris disks similar to those observed around nearby main sequence stars 
\citep{wya08}.

Numerical simulations of icy planet formation suggest the cascade limits the 
masses of growing protoplanets to $\sim$ 0.01 \mearth\ \citep[][hereafter KB08]
{kb08}. This mass is much smaller than the core mass, $\gtrsim$ 0.1--1 \mearth, 
required for a protoplanet to accrete gas and become a gas giant planet
\citep{pol96,ina03,ali05}.  Unless icy protoplanets can accrete collision 
fragments before the fragments are ground to dust, these protoplanets cannot 
grow into gas giant planet cores.  Thus, finding a mechanism to halt the cascade 
is essential to form gas giant planets.  

Here, we describe how interactions between the fragments and the gaseous 
disk can halt the cascade. In our picture, the gas traps small fragments with 
sizes of 0.1~mm to 1~m and prevents them from colliding at large velocities. 
These fragments then settle rapidly to the disk midplane, where protoplanets 
can accrete them. For a broad range of initial conditions, analytic results 
and detailed numerical simulations demonstrate that this process yields 
1--10~\mearth\ cores in 1--2~Myr.

We develop the analytic theory in \S2 and derive the conditions needed for
protoplanets to accrete collision fragments and grow to masses of $\sim$ 
1 \mearth\ in 1--2~Myr. We confirm these estimates in \S3 with detailed 
numerical calculations. We conclude with a brief discussion in \S4.

\section{PHYSICAL MODEL}

The crucial element of our model is the interaction of collision fragments with 
the gaseous disk. Fragments larger than the `stopping radius' $r_s \approx$ 
0.5--2~m at 5--10 AU \citep{wei77,raf04}, orbit with the growing protoplanets, 
independently of the gas. Destructive collisions among these fragments fuel the 
collisional cascade. However, the gas entrains particles with radii $r \lesssim r_s$.  
These fragments orbit with the gas; thus, their velocity dispersions are small 
and independent of massive protoplanets. By trapping small collision fragments, 
the gas halts the collisional cascade.

The gas also allows protoplanets to accrete the debris. When the collisional 
cascade begins, the mass in leftover planetesimals is $\sim$ 1--10~\mearth. 
The cascade grinds all of this mass into small fragments which are trapped by 
the gas.  Most of the trapped fragments fall through the gas into the midplane 
of the disk, where growing protoplanets accrete them.  Protoplanets that 
accrete $\sim$ 0.1--1~\mearth\ before the gas dissipates 
\citep[$\sim$ 3--10~Myr;][]{hart98,hai01,kenn09} become gas giants.  Thus, 
our model yields gas giant cores if (i) the collisional cascade produces 
fragments fast enough, (ii) the fragments quickly settle to the midplane, 
and (iii) the largest protoplanets rapidly accrete the fragments.

To examine whether this physical model leads to cores with masses of 
$\sim$ 1 \mearth, we consider the growth of planets in a disk of gas 
and icy objects around a star with mass \mstar.  Material at a distance 
$a$ from the central star orbits with angular frequency $\Omega$ and has 
surface densities $\Sigma_s$ (solids) and $\Sigma_g$ (gas).  We adopt a 
solid-to-gas ratio of 1:100 and $\Sigma_s = \Sigma_{s,0} ~ x_m ~ a^{-3/2}$, 
where $\Sigma_{s,0}$ = 2.5 g cm$^{-2}$ at 5 AU and $x_m$ is a scale factor.

Forming icy protoplanets is the first step in our model.  In an ensemble of 
1~km planetesimals, collisional growth yields a few 1000~km objects -- 
`oligarchs' -- that contain an ever-increasing fraction of the mass in 
solids \citep{ida93,wet93,raf03}.  From numerical simulations of 
planet growth at 30--150 AU, the timescale to produce an oligarch around
a solar-type star is (KB08)
\begin{equation}
t_{1000} \sim 10^5 ~ x_m^{-1.15} ~ \left ( \frac{a}{\rm 5~AU} \right )^3 ~
{\rm yr} ~ .
\label{eq:t1000allm}
\end{equation}
Thus, oligarchs form at 5 AU before the gas dissipates.

Once oligarchs form, collisions among leftover planetesimals produce copious 
amounts of fragments.  In the high velocity limit, the collision time for a 
planetesimal of mass $M$ in a swarm of icy planetesimals with mass $M$, radius 
$r$, and surface density $\Sigma$ is $t_c$ 
$\approx M / (\Sigma~\pi~r^2~\Omega)$ \citep{gol04}.  Thus,
\begin{equation}
t_c \approx 10^5 ~ x_m^{-1} \left ( \frac{r}{\rm 1 ~ km} \right ) \left ( \frac{a}{\rm 5 ~ AU} \right )^{3/2} ~ {\rm yr} ~ .
\label{eq:tcoll}
\end{equation}
Collisions among planetesimals produce debris at a rate 
$\dot{M} \approx N ~ \delta M ~ t_c^{-1}$, where $N$ is 
the number of planetesimals of mass $M$ and $\delta M$ is 
the mass in fragments produced in a single collision.  In an 
annulus of width $\delta a$ at distance $a$ from the central star,
$N \approx 2 \pi ~ \Sigma ~a ~\delta a / M$. If $\sim$ 10\% of
the mass in each collision is converted into fragments
\begin{equation}
\dot{M_f} \approx 4 \times 10^{-7} ~ x_m^2 ~ \left ( \frac{\rm 5~AU}{a} \right )^{7/2} \left ( \frac{\delta a}{\rm 0.2~AU} \right ) \left ( \frac{\rm 1~km}{r} \right) ~ M_{\oplus}~{\rm yr^{-1}} ~ ,
\label{eq:mdotfrag}
\end{equation}
where we have set the width of the annulus equal to the width of the
`feeding zone' for a 0.1~\mearth\ protoplanet \citep{lis87}.
Thus, disks with $x_m \gtrsim$ 1--2 produce fragments at a rate sufficient
to form $\gtrsim$ 1~\mearth\ cores in 1--2~Myr.

Most of the mass in fragments settles quickly to the disk midplane.  For a 
settling time $t_s \approx 4 x_m ({\rm 1~m}/r)$ yr \citep{chi97}, fragments 
with $r \gtrsim$ 0.1 mm reach the midplane on the collision timescale of 
$\sim 10^5$~yr (Eq. \ref{eq:tcoll}). For a size distribution $n(m)$
$\propto m^{q}$ with $q$ = $-1$ to $-0.8$ \citep{doh69,hol07}, 66\% to 99\% 
of the total mass in fragments with $r \lesssim$ 1~m settles to the 
midplane in $\lesssim 10^5$ yr at 5--10 AU.  

Oligarchs rapidly accrete fragments in the midplane. The maximum accretion 
rate for an oligarch with $M_o \sim$ 0.01~\mearth\ at 5~AU is 
$\sim 5 \times 10^{-6}$  \mearth\ yr$^{-1}$ \citep{raf04}. This maximum rate 
yields 5 \mearth\ cores in 1~Myr.  At the onset of the cascade, 
our simulations suggests oligarchs at 5 AU accrete at a rate 
\begin{equation}
\dot{M_o} \approx 10^{-6} \left ( \frac{M_o}{0.01~M_{\oplus}} \right )^{2/3} ~ \left ( \frac{M_f}{1~M_{\oplus}} \right ) ~ M_{\oplus}~{\rm yr^{-1}} ~ ,
\label{eq:mdotoli}
\end{equation}
where $M_f$ is the total mass in fragments in a feeding zone with width
$\delta a \approx$ 0.2 AU at 5 AU. Thus, protoplanets likely reach masses 
$\sim$ 1~\mearth\ in 1--2~Myr.

These analytic estimates confirm the basic aspects of our model.  In a 
gaseous disk with $\Sigma_g \approx$ 250 g~cm$^{-2}$, the gas halts the 
collisional cascade. Collision fragments entrained by the gas rapidly settle
to the midplane.  Protoplanets with masses $\sim$ 0.01 \mearth\ can accrete
collision fragments rapidly and grow to masses $\sim$ 1 \mearth\ before the
gas dissipates in 3--10~Myr.

\section{NUMERICAL MODEL}

To explore this picture in more detail, we calculate the formation of 
cores with our hybrid multiannulus coagulation--$n$-body code \citep{bk06}. 
In previous calculations, we followed the evolution of objects with 
$r \gtrsim r_s$; collision fragments with $r \lesssim r_s$ were removed by the 
collisional cascade (KB08). Here, we include the evolution of small fragments 
entrained by the gas. We follow \citet{bra08a} and calculate the scale height of 
small particles with $r < r_s$ as $H = \alpha H_g / [{\rm min}(St,0.5)(1 + St)]$, 
where $H_g$ is the scale height of the gas \citep[][KB08]{kh87}, $\alpha$ is the 
turbulent viscosity of the gas, and $St = r \rho_s \Omega / c_s \rho_g$.  In this 
expression for the Stokes number ($St$), $c_s$ is the sound speed of the gas, 
$\rho_g$ is the gas density, and $\rho_s$ is the mass density of a fragment.
We assume small particles have vertical velocity $v = H \Omega$ and horizontal 
velocity $h = 1.6 v$. Protoplanets accrete fragments at a rate $n \sigma v_{rel}$,
where $n$ is the number density of fragments, $\sigma$ is the cross-section
(including gravitational focusing), and $v_{rel}$ is the relative velocity
\citep[e.g.,][Appendix A.2]{kl98}. Although this approximation neglects many 
details of the motion of particles in the gas \citep{bra08a}, it approximates 
the dynamics and structure of the fragments reasonably well and allows us to 
calculate accretion of fragments by much larger oligarchs. 

Using the statistical approach of \citet{saf69}, we evolve the masses 
and orbits of planetesimals in a set of concentric annuli with widths 
$\delta a_i$ at distances $a_i$ from the central star (KB08). The 
calculations use realistic cross-sections (including gravitational 
focusing) to derive collision rates \citep{spa91} and a Fokker-Planck 
algorithm to derive gravitational stirring rates \citep{oht02}. When 
large objects reach a mass $M_{pro}$, we `promote' them into an $n$-body 
code \citep{bk06}. This code follows the trajectories of individual 
objects and includes algorithms to allow interactions between the massive 
$n$-bodies and less massive objects in the coagulation code.

To assign collision outcomes, we use the ratio of the center of mass 
collision energy $Q_c$ and the energy needed to eject half the mass 
of a pair of colliding planetesimals to infinity $Q_d^*$. We adopt
$Q_d^* = Q_b r^{\beta_b} + Q_g \rho_g r^{\beta_g}$ \citep{ben99},
where $Q_b r^{\beta_b}$ is the bulk component of the binding energy,
$Q_g \rho_g r^{\beta_g}$ is the gravity component of the binding energy,
and $\rho_g$ is the mass density of a planetesimal. The mass of a merged 
pair is $M_1 + M_2 - M_{ej}$, where the mass ejected in the collision 
is $M_{ej} = 0.5 (M_1 + M_2) (Q_c/Q_d^*)^{9/8}$ \citep{kl99}. 

Consistent with recent $n$-body simulations, we consider two sets of 
fragmentation parameters $f_i$. Strong planetesimals have $f_S$ = 
\citep[$Q_b$ = 1, $10^3$, or $10^5$ erg g$^{-1}$, $\beta_b \approx$ 0, 
$Q_g$ = 1.5 erg g$^{-1}$ cm$^{-1.25}$, $\beta_g$ = 1.25; KB08, ][]{ben99}. 
Weaker planetesimals have $f_W$ = 
\citep[$Q_b$ = $2 \times 10^5$ erg g$^{-1}$ cm$^{0.4}$, $\beta_b \approx -0.4$, 
$Q_g$ = 0.22 erg g$^{-1}$ cm$^{-1.3}$, $\beta_g$ = 1.3;][]{lei08}.

Our initial conditions are appropriate for a disk around a young star 
\citep[e.g.][]{dul05,cie07a,gar07,bra08b}.  We consider systems of 32 annuli 
with $a_i$ = 5--10 AU and $\delta a_i/a_i$ = 0.025. The disk is composed 
of small planetesimals with radii ranging from $r_{min} = r_s \approx$ 
0.5--2~m \citep{raf04} to $r_0$ = 1~km, 10~km, or 100~km and an initial mass 
distribution $n_i(M_{ik}) \propto M_{ik}^{-0.17}$. The mass ratio between 
adjacent bins is $\delta = M_{ik+1}/M_{ik}$ = 1.4--2 \citep[e.g.,][KB08]{kl98}.  
Each bin has the same initial eccentricity $e_0 = 10^{-4}$ and inclination 
$i_0 = e_0/2$. 

For each combination of $r_0$, $f_i$, and $x_m$ = 1--5, we calculate 
the growth of oligarchs with two different approaches to grain accretion.
In models without grain accretion, fragments with $r \lesssim r_{min}$ 
are `lost' to the grid. Oligarchs cannot accrete these fragments; their 
masses stall at $M \lesssim$ 0.1~\mearth.  In models with grain accretion, 
we track the abundances of fragments with 0.1~mm $ \lesssim r \lesssim r_{min}$ 
which settle to the disk midplane on short timescales.  Oligarchs can accrete 
these fragments; they grow rapidly at rates set by the production of collision 
fragments.

For the gaseous disk, we adopt $\alpha = 10^{-4}$, an initial surface density, 
$\Sigma_{g,0}$ = 100~$\Sigma_{s,0}~x_m~a^{-3/2}$, and a depletion time $t_g$ = 3~Myr. 
The surface density at later times is $\Sigma_{g,t} = \Sigma_{g,0}~e^{-t/t_g}$.
We ignore the migration of protoplanets from torques between the gas and 
the planet \citep{lin86,war97}. \citet{ali05} show that migration enhances 
growth of protoplanets; thus our approach underestimates the growth time. 
We also ignore the radial drift of fragments coupled to the gas. Depending
on the internal structure of the disk, fragments can drift inward, drift 
outward, or become concentrated within local pressure maxima or turbulent
eddies \citep{wei77,hagh2003,inaba06,masset06,cie07b,kret07,kato08}.  Here, 
our goal is to provide a reasonable first estimate for the growth rates 
of protoplanets.  We plan to explore the consequences of radial drift in 
subsequent papers.  

\section{RESULTS}

\begin{figure}
\includegraphics[height=5.0in]{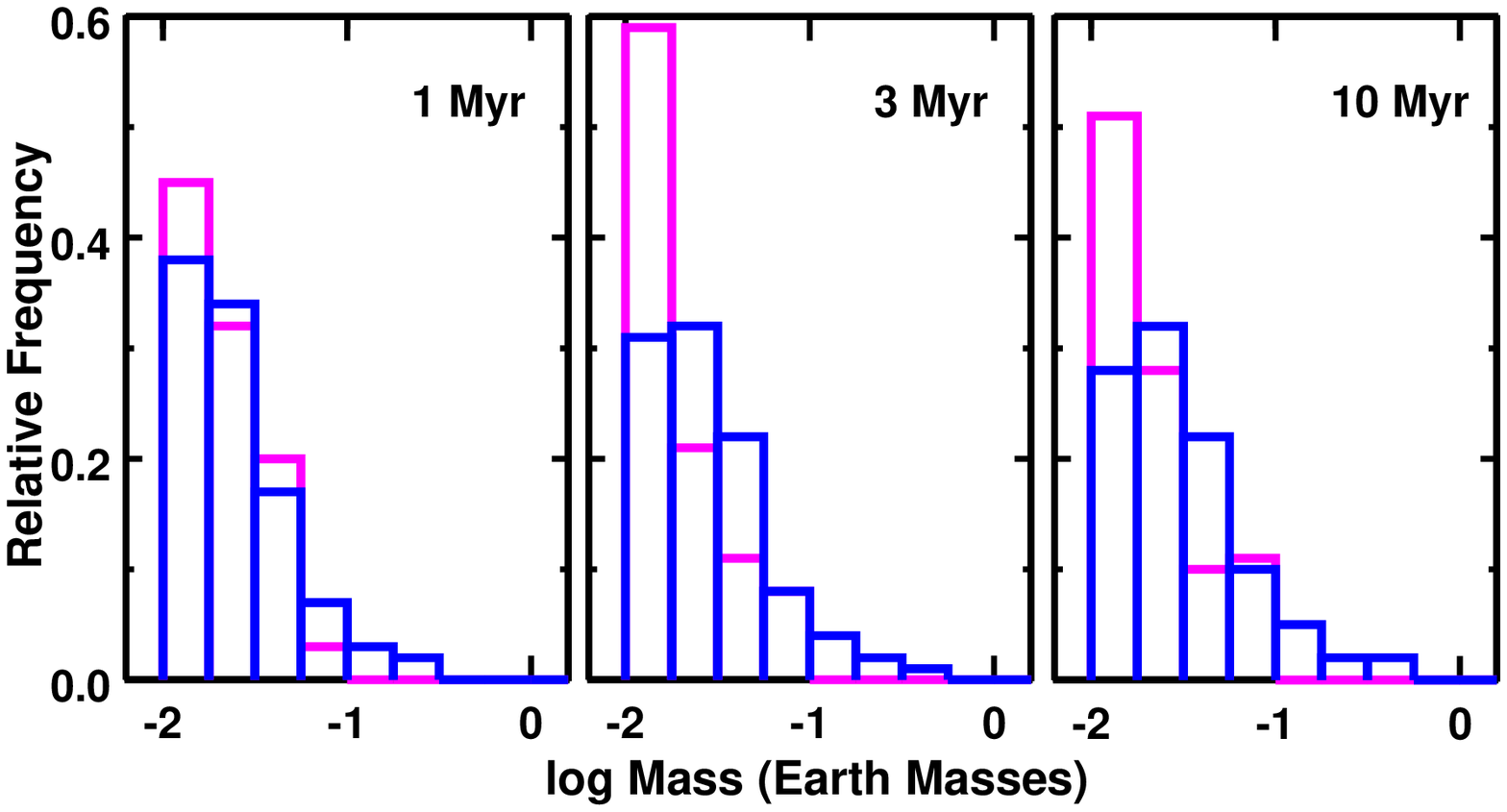}
\figcaption[f1.eps]
{
Mass histograms at 1~Myr (left panel), 3~Myr (center panel), and
10~Myr (right panel) for coagulation calculations without grain 
accretion using the $f_S$ fragmentation parameters at 5 AU.  Magenta 
histograms plot median results for 25 calculations with $x_m$ = 1; 
blue histograms show median results for 25 calculations with $x_m$ 
= 5.  Independent of disk mass, calculations without grain accretion
yield planets with maximum masses $\lesssim$ 1 \mearth\ in 10~Myr.
\label{fig:sd1}
}
\end{figure}

Fig. \ref{fig:sd1} shows mass histograms at 1--10~Myr for coagulation 
calculations without grain accretion using $r_0$ = 1 km and the strong 
fragmentation parameters ($f_S$). After the first oligarchs with $M \sim$ 
0.01 \mearth\ form at $\sim$ 0.1~Myr, the collisional cascade starts 
to remove leftover planetesimals from the grid. Independent of $Q_b$, 
the cascade removes $\sim$ 50\% of the initial mass of the grid in 
$\sim$ 4 $x_m^{-1.25}$~Myr. As the cascade proceeds, growth of the
largest oligarchs stalls at a maximum mass $M_{o,max} \approx$ 
0.1 $x_m$ \mearth. 

These results depend weakly on $r_0$.  The time to produce the first 
oligarch with $r \sim$ 1000~km increases with $r_0$, 
$t_{1000} \sim 0.1~x_m^{-1.25}~(r_0/{\rm 10~km})^{1/2}$~Myr. Calculations 
with larger $r_0$ tend to produce larger oligarchs at 10~Myr:
$M_{o,max} \approx$ 1 \mearth\ (2 \mearth) for $r_0$ = 10~km (100~km).
In $\sim$ 50 calculations, none produce cores with $M_{o,max} \gtrsim$ 
1 \mearth\ on timescales of $\lesssim$ 10~Myr.

For $r_0 \lesssim$ 100~km, our results depend on $f_i$.  In models with
$r_0$ = 1~km and 10~km, the $f_W$ fragmentation parameters yield oligarchs 
with smaller maximum masses, $M_{o,max} \approx$ 0.3--0.6~\mearth. Because
leftover planetesimals with $r \sim$ 1--10~km fragment more easily, the 
cascade begins (and growth stalls) at smaller collision velocities when 
oligarchs are less massive \citep{kbod08}. 

\begin{figure}
\includegraphics[height=5.0in]{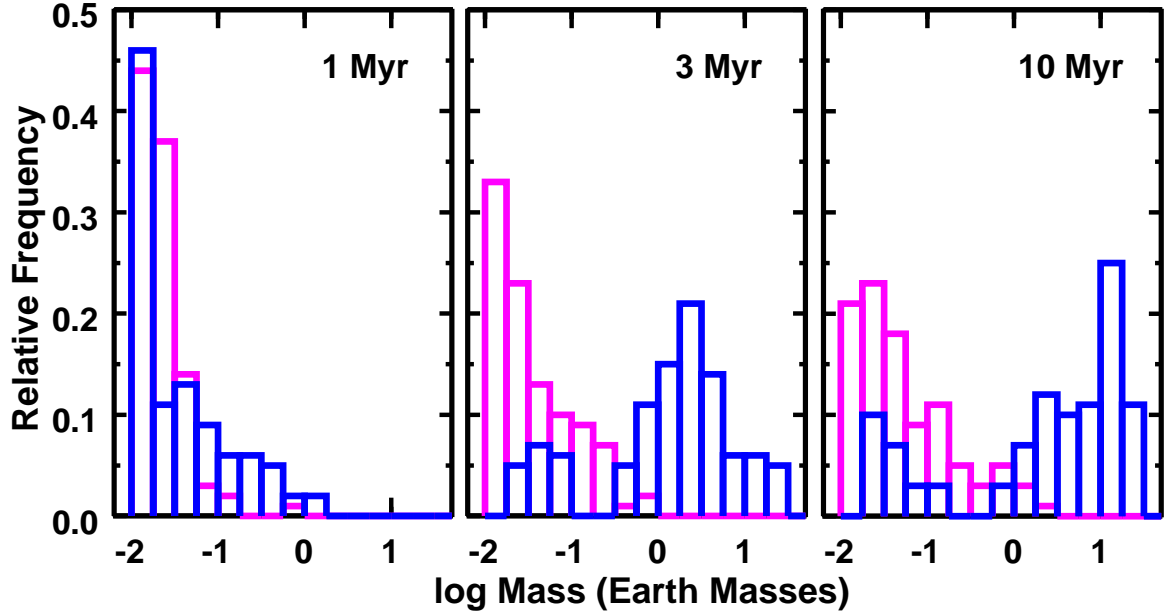}
\figcaption[f2.eps]
{
As in Fig. 1 for calculations with grain accretion.  When 
large oligarchs can accrete fragments trapped by the gas, 
disks with $x_m \gtrsim$ 1 produce gas giant cores in 3--10~Myr.
\label{fig:sd2}
}
\end{figure}

Calculations with grain accretion produce cores rapidly.  Fig. \ref{fig:sd2} 
shows results at 1--10~Myr for calculations with $r_0$ = 1~km and the $f_S$ 
fragmentation parameters. As the first oligarchs reach masses of $\sim$ 
0.01 \mearth\ at 0.1~Myr, the cascade generates many small collision 
fragments with $r \sim$ 1~mm to 1~m. These fragments rapidly settle to 
the disk midplane and grow to sizes of 0.1--1~m. When the cascade has 
shattered $\sim$ 25\% of the leftover planetesimals, oligarchs begin a 
second phase of runaway growth by rapidly accreting small particles in 
the midplane. For calculations with $x_m$ = 1--5, it takes $\sim$ 
1--2~$x_m^{-1.25}$~Myr to produce at least one core with $M_o \gtrsim$ 
1--5~\mearth. Thus, cores form before the gas dissipates. 

These results depend on $r_0$. For $r_0$ = 10~km, fragmentation produces 
small grains 2--3 times more slowly than calculations with $r_0$ = 1~km. 
These models form cores more slowly, in 5--10 $x_m^{-1.25}$~Myr instead 
of 1--2 $x_m^{-1.25}$~Myr. For models with $r_0$ = 100~km, fragmentation 
yields a negligible mass in small grains.  Thus, cores never form in 
$\lesssim$ 10--20~Myr.  

The timescales to form cores also depend on $f_i$.  Calculations with the 
$f_W$ parameters form cores 10\% to 20\% faster than models with the 
$f_S$ parameters. 

\section{CONCLUSIONS}

Gaseous disks are a crucial element in the formation of the cores of
gas giant planets. The gas traps small collision fragments and halts 
the collisional cascade. Once fragments settle to the disk midplane,
oligarchs accrete the fragments and grow to masses $\gtrsim$
1 \mearth\ in 1--3~Myr.

Our model predicts two outcomes for icy planet formation. Oligarchs that 
form before (after) the gas disk dissipates reach maximum masses $\gtrsim$ 
1 \mearth\ ($\lesssim$ 0.01--0.1 \mearth). Setting the timescale to form
a 1000~km oligarch (Eq. \ref{eq:t1000allm}) equal to the gas dissipation
timescale $t_g$ yields a boundary between these two types of icy protoplanets 
at $a_g \approx$ 15 $x_m^{0.4} (t_g / {\rm 3~Myr})^{1/3}$~AU.  We expect 
massive cores at $a \lesssim a_g$ and low mass icy protoplanets at $a \gtrsim a_g$.

This prediction has a clear application to the Solar System.  Recent dynamical 
calculations suggest that the Solar System formed with four gas giants at 
5--15~AU and an ensemble of Pluto-mass and smaller objects beyond 20~AU 
\citep{mor08}. For a protosolar disk with $x_m^{1.2}~(t_g / {\rm 3~Myr})$
$\approx$ 1, our model explains this configuration. Disks with these
parameters are also common in nearby star-forming regions \citep{and05}. 
Thus, our results imply planetary systems like our own are common.

Our model yields a large range of final masses for massive icy cores.
Protoplanets that grow to a few \mearth\ well before the gas dissipates 
can accrete large amounts of gas from the disk and become gas giants 
\citep{pol96,ali05}.  Protoplanets that grow more slowly cannot accrete 
much gas and become icy `super-Earths' with much lower masses 
\citep{kenn06,kenn08}. For solar-type stars with $t_g \approx$ 3~Myr, 
our results suggest that gas giants (super-Earths) are more likely in 
disks with $x_m \gtrsim$ 1.5 ($x_m \lesssim$ 1.5) at 5--10 AU.

Testing this prediction requires (i) extending our theory to a range of
stellar masses and (ii) more detections of massive planets around lower 
mass stars.  We plan to explore the consequences of our model for other 
stellar masses in future papers. Larger samples of planetary systems will 
test the apparent trend that gas giants (super-Earths) are much more common 
around solar-type (lower mass) stars \citep[e.g.,][]{cum08,for08}. 
Comparing the results of our planned numerical calculations with these 
additional observations will yield a clear test of our model.

\vskip 6ex

We acknowledge a generous allotment, $\sim$ 25 cpu years, of computer
time on the 1024 cpu Dell Xeon cluster `cosmos' at the Jet Propulsion
Laboratory through funding from the NASA Offices of Mission to Planet
Earth, Aeronautics, and Space Science.	We thank M. Werner for his strong
support of this project.  We also acknowledge use of $\sim$ 10 cpu years
on the CfA cluster `hydra.' Advice and comments from T. Currie, M. Geller,
G. Kennedy, and an anonymous referee greatly improved our presentation.  
Portions of this project were supported by the {\it NASA } 
{\it TPF Foundation Science Program,} through grant NNG06GH25G.

{}

\clearpage

\end{document}